\documentstyle[12pt,epsfig]{article}
\newlength{\dinwidth}
\newlength{\dinmargin}
\newcommand{\ba}{\begin{array}}
\newcommand{\ea}{\end{array}}
\newcommand{\be}{\begin{equation}}
\newcommand{\ee}{\end{equation}}
\newcommand{\bea}{\begin{eqnarray}}
\newcommand{\eea}{\end{eqnarray}}
\newcommand{\gsim}{\mathrel{\mathop{\kern 0pt \rlap
  {\raise.2ex\hbox{$>$}}} \lower.9ex\hbox{\kern-.190em $\sim$}}}

\def\ben{\begin{equation}}
\def\een{\end{equation}}
\def\bea{\begin{eqnarray}}
\def\eea{\end{eqnarray}}

\begin{document}
\thispagestyle{empty}
\addtocounter{page}{-1}
\vskip-0.35cm
\begin{flushright}

IFT UAM/CSIC/02-53\\
{\tt hep-th/0211137}
\end{flushright}
\vspace*{0.2cm}
\centerline{\Large \bf A Comment on Masses, Quantum Affine Symmetries}
\vskip0.3cm
\centerline{\Large \bf and 
PP-Wave Backgrounds}
\vspace*{1.0cm} 
\centerline{\bf Cesar Gomez }
\vspace*{0.7cm}
\centerline{\it Instituto de Fisica Teorica, C-XVI Universidad Autonoma,}
\vspace*{0.2cm}
\centerline{\it E-28049 Madrid \rm SPAIN}
\vspace*{1cm}
\centerline{\tt Cesar.Gomez@uam.es}

\vspace*{0.8cm}
\centerline{\bf abstract}
\vspace*{0.3cm}
Two dimensional light cone world sheet massive models can be used to
define good string backgrounds.In many cases these light cone world sheet
lagrangians flow from a CFT in the UV to a theory of massive particles
in the IR. The relevant symmetry in the IR, playing a similar role to
Virasoro in the UV, are quantum affine Kac Moody algebras. Finite
dimensional irreps of this algebra are associated with the
spectrum of massive particles. The
case of $N=0$ Sine Gordon at the $N=2$ point is associated
with a Landau Ginzburg model that defines a good string background. 
For the world sheet symmetry $(N=2) \otimes U_{q}(\hat{Sl(2)})$ the 
$N=2$ piece is associated with the string conformal invariance and
the $U_{q}(\hat{Sl(2)})$ piece with the world sheet RG.
The two dimensional
light cone world sheet massive model can be promoted to a CFT by adding
extra light cone fields $X^{-}$ and $X^{+}$. From the point of view
of the quantum affine symmetry these two fields are associated, respectively, 
with the
center and the derivation of the affine Kac Moody algebra.

\baselineskip=18pt
\newpage

\section{Introduction}
Two dimensional conformal field theories are, from the world sheet point of 
view, the building blocks of string theories. Massive two dimensional models
that don't contain derivative couplings are superrenormalizable
and asymptotically free and therefore they flow in the UV to conformal 
field theories. Some massive models are specially simple
namely, integrable models. These models generically enjoy quantum affine 
symmetries \cite{BLC} with the soliton S matrix being given in terms of the 
quantum R matrix. A classical example is the Sine Gordon model.

Recently and motivated by the study of type IIB superstring in the
Penrose limit \cite{P} \cite{blau} of $AdS_{5}\times S^{5}$ \cite{metsaev}
\cite{BMN} some new pp-wave metrics with RR backgrounds 
have been discovered \cite{MN}.
The interesting thing about these backgrounds is that they are 
naturally associated with massive two dimensional models. These massive
models are just the light cone gauge world sheet lagrangian in these
backgrounds.

In the standard flat background case, the light cone world sheet
lagrangian for the transversal coordinates is conformal invariant and
we can easily define the corresponding Virasoro algebra with
central extension $D-2$ and the Virasoro generators $L_{n}$, $\bar L_{n}$ 
defined
in terms of the transversal oscillators. In pp wave backgrounds we generically
find massive deformations of this light cone world sheet lagrangian.
In some special cases these massive models possess quantum affine symmetries
(in the infinite volume limit) of type
$U_{q}(\hat{Sl(2)})$ at level $k=0$ \cite{BLC}. 
This quantum affine symmetry replace the
Virasoro symmetry of the flat case. 

Longitudinal boosts of the target metric define the RG of 
the massive world sheet light cone lagrangian \cite{MN}. 
In the UV the system flow to the 
conformal model and in the IR the massive particle spectrum dominates. 
This particle spectrum is generically 
associated with the finite dimensional irreps
of the quantum affine symmetry at level $k=0$ \cite{SLC}.

\section{Massive light cone lagrangians and good string backgrounds}

Given a two dimensional superrenormalizable massive model
of type 
\ben\label{L}
S = \int d^{2} \sigma ( \partial X^{i} \partial X^{i} + V(X^{i}))
\een
we can always define, classically, a conformal invariant world sheet
action just adding extra world sheet fields $X^{+}$ and $X^{-}$
as follows
\ben\label{M}
S= \int d^{2} \sigma \sqrt{h} h_{a,b} (\partial_{a}X^{i} \partial_{b} X^{i}
+ \partial_{a} X^{+} \partial_{b} X^{-} - 
V(X) \partial_{a}X^{+} \partial_{b}X^{+})
\een
Since
\ben
\partial_{+} \partial_{-} X^{+} = 0
\een
we can interpret the original massive model (\ref{L}) as the result of fixing
the light cone gauge, and the variables $X^{i}$ as representing the transversal
coordinates \footnote{ Notice that the addition of a 
total derivative to lagrangian (\ref{L}) is equivalent to add a Kalb Ramond 
background $B$ with non vanishing components $B_{+ i}$} .

Of course the world sheet model (\ref{M}) will not define, in general, a good
string background. The contraction of $X^{i}$ in
(\ref{M}) induce non vanishing sigma model beta functions $\beta_{G}$.
The simplest example corresponding to the maximally symmetric 
pp-wave where $V(X^{i}) = 
\mu X^{i} X^{i}$ produces, at one loop,  

a non vanishing sigma model beta function
\ben
\beta_{G} = R_{+ +} = 8\mu
\een
which is, of course, the curvature of the maximally symmetric
pp-wave metric
\ben
ds^{2} = dx^{+}dx^{-} + \mu x^{i} x^{i}dx^{+} dx^{+} + dx^{i}dx^{i}
\een

In order to get a good string background from the massive model
(\ref{L}) with $V(X^{i}) = 
\mu X^{i} X^{i}$  we need to add a RR background. We can in principle
get the RR background solving the
type IIB supergravity equations for a RR five form
\ben
F_{5} = dX^{+} \wedge \phi(X^{i})
\een

After adding this RR background the corresponding light cone world sheet
lagrangian can be writen as a $N=2$ lagrangian
\ben\label{R}
S = \int d^{2}z d^{4} \theta K(\Phi \bar\Phi) + 
\int d^{2}z d^{2} \theta W(\Phi) + c.c
\een
with
\ben\label{s}
K(\Phi^{i}, \bar \Phi^{i}) = \Phi^{i} \bar \Phi^{i}
\een
and superpotential, in complex coordinates,
\ben
W(\Phi)=\frac{\mu}{2}\sum_{l}(\Phi^{l})^{2}
\een
for $l=1...4$.

In particular it follows from Berkovits formalism \cite{B} \cite{BM} 
that any $N=2$
sigma model (\ref{R}) with flat transversal manifold, i.e $K$ as defined
in (\ref{s}), and arbitrary holomorphic superpotential $W$ defines a good 
pp-wave string background
\ben
ds^{2} = dx^{+}dx^{-} + H(x^{i}) dx^{+}dx^{+} + dz^{i}dz^{i}
\een
with
\ben
H = \partial W \bar\partial \bar W
\een
and the RR form a $(1,3)$ and $(3,1)$ form \cite{MN}.

\subsection{Landau-Ginsburg  string backgrounds}
According with the previous discussion we can define pp-wave string 
backgrounds associated with light cone world sheet 
lagrangians of the type of $N=2$ LG models \cite{VW} \cite{Mar}. 
We will consider the 
universality class defined by the Kahler potential (\ref{s}). The maximally
symmetric pp-wave background corresponds in this language
to the universality class of the massive free superfield. The associated
chiral ring is trivial, generated by the identity, and the associated
central charge is $c=0$ \footnote{ As a marginal comment
let us just notice that Lipatov model for N=0 Yang Mills
is a XXX spin chain with spin $j=0$ \cite{KF}. 
Naively the central extension for this 
model
will be also $c=0$}.

The non renormalization theorems for $N=2$ allow us to associate with a LG
lagrangian with a quasihomogenous
superpotential a $N=2$ SCFT as the fixed point in the IR. The chiral ring
of this $N=2$ SCFT is the one determined by the superpotential $W$.
Given the relation between $W$ and the RR background we can interpret the
RR background as inducing the massive deformation of the light cone
world sheet lagrangian.

In the case of the maximally symmetric pp-wave we find 
a flow, for the world sheet RG induced by longitudinal boosts, from
$c=1$\footnote{ We normalize the central extension of the
free $N=2$ superfield to $1$ instead of the usual $c=3$} in the UV to
the trivial fixed point in the IR corresponding to $c=0$. This flow is
of course consistent with Zamolodchikov c-theorem $c_{UV}>c_{IR}$.
The $c=0$ fixed point in the IR 
corresponds to the Kac Moody algebra $\hat{Sl(2)}$ at level
$k=0$. As it is well known this algebra admits finite dimensional irreps
that we should associate with a particle like spectrum in the IR.
The maximally symmetric pp wave $V(X^{i}) = \mu X^{i}X^{i}$ in the deep IR 
corresponding to $\mu \rightarrow \infty$ was considered in reference
\cite{FF}. In this limit the string reduces to a set independent bits 
characterized in terms of two dimensional irreps.

\section{Quantum Affine Symmetry of $N=0$ Sine Gordon}
The $N=0$ SG theory is defined by the lagrangian
\ben
S= \frac{1}{4 \pi}\int d^{2}z (\partial_{z}\Phi \partial_{\bar z} \Phi 
+ \frac{\lambda}{\pi}
:cos( \beta \Phi):)
\een
This theory for
$0< \beta^{2} <2$ 
defines a deformation of the conformal field theory with $c=1$
of a single scalar field. In the UV, corresponding to $\lambda =0$
the theory flow to this conformal field theory. It is well known that
this theory possess a quantum affine symmetry $U_{q}(\hat{Sl(2)})$ \cite{BLC}.
This is the quantum deformation of the affine Kac Moody algebra
$\hat{Sl(2)}$ at level $k=0$. The value of $q$
is given in terms of $\beta$ as follows
\ben\label{O}
q= exp( \frac{-2\pi i}{{\beta}^{2}})
\een
At special values of the coupling
\ben
\beta = \sqrt{ \frac{2p}{p+1}}
\een
we get $q= - exp( \frac{-i \pi}{p})$ and therefore the $N=2$
superalgebra for the special value $p=2$.
The existence of this quantum symmetry is proved in the infinite volume limit.
Let us briefly review the way this symmetry is defined.
The first step is to define, associated with the SG field $\Phi$, 
the equivalent to chiral and antichiral pieces $\Phi = \phi + \bar \phi$.
This is done introducing non local fields
\ben
\phi(\sigma, t) =\frac{1}{2}(\Phi(\sigma, t) + 
\int_{- \infty}^{\sigma}dy \partial_{t} \Phi(y,t))
\een
\ben
\bar \phi(\sigma, t) =\frac{1}{2}(\Phi(\sigma, t) - 
\int_{- \infty}^{\sigma}dy \partial_{t} \Phi(y,t))
\een
Using Zamolodchikov formalism \cite{Z} \cite{EY} 
we can now define conserved currents
as
\ben
J_{\pm}= exp(\pm \frac{2i}{\beta} \phi)
\een
\ben
\bar J_{\pm}= exp(\mp \frac{2i}{\beta} \bar \phi)
\een

The corresponding charges generate the affine quantum algebra 
$U_{q}(\hat{Sl(2)})$
\ben
Q_{+}\bar Q_{+} -q^{2}\bar Q_{+}Q_{+} = 0
\een
\ben
Q_{-} \bar Q_{-} - q^{2} \bar Q_{-} Q_{-} = 0
\een
\ben
Q_{+} \bar Q_{-} - q^{-2} \bar Q_{-} Q_{+} = a( 1- q^{2T})
\een
\ben
Q_{-} \bar Q_{+} - q^{-2} \bar Q_{+} Q_{-} = a( 1- q^{- 2T})
\een
with
\ben
a= \frac{\lambda \gamma^{2}}{2\pi i}
\een
with 
\ben
e^{- \frac{2\pi i}{\beta^{2}}} = - e^{- \frac{i \pi}{\gamma}}
\een
and  $T$ , the topological charge defined as
\ben
T = \frac{\beta}{2 \pi} \int_{- \infty}^{+ \infty} \partial_{x} \Phi
\een
Notice that at the particular point $p=2$ we recover the $N=2$ algebra.
The $N=0$ SG model at this particular point 
is equivalent to the following $N=2$ model \cite{LCV}
\ben\label{N}
S = \int d^{2}z d^{4} \theta X X^{*} + (\int d^{2}z d^{2} \theta 
\lambda ( \frac{1}{3}X^{3} - X) +c.c)
\een
Interpreted as a Landau Ginsburg lagrangian the case $p=2$
correspond to a chiral ring with two elements $1,X$. The associated
central extension is $c=1$. 
In this sense it looks that the $N=0$ 
SG at the $N=2$ point is just another particular case of a pp-wave background. 
The potential interest of this case is that we start with a $N=0$
theory that at some particular value of the coupling becomes $N=2$ and
therefore equivalent to a $N=2$ LG model that defines a good string background.

\section{Strings and Quantum Affine Symmetries}
Let us consider the light cone lagrangian in flat space time. For this two
dimensional model we have a set of Noether conserved currents that correspond
to isometries of the target space time. Among these symmetries the ones
associated with currents $J^{i -}$ and $J^{+ -}$ describe, in particular, the 
scale (conformal) invariance of the two dimensional 
light cone world sheet model. From 
target space time point of view they are part of Lorentz invariance. Target
space time metrics that break Lorentz, as it is the case of pp-waves, produce
massive world sheet two dimensional models for which the currents 
$J^{i -}$ and $J^{+ -}$ are not conserved. However these massive models
can still have symmetries associated with non local currents. Symmetries 
that replace in the massive case conformal invariance. These symmetries
are, as we have review in previous section, quantum deformations of affine
Kac Moody algebras at level zero. Finite dimensional
representations of these algebras as well as quantum R-matrices characterize
the spectrum of massive particles \footnote{ Finite
dimensional irreps of $U_{q}(\hat{Sl(2)})$ are defined as follows:
\ben
Q_{\pm} = e^{\frac{2 \theta}{\beta^{2}}} E_{\pm} q^{\pm \frac{H}{2}}
\een
\ben
\bar Q_{\pm} = e^{- \frac{2 \theta}{\beta^{2}}} E_{\pm} q^{\mp \frac{H}{2}}
\een
where $q= e^{\frac{-2 \pi i}{\beta^{2}}}$ , $E_{\pm}$ are the Pauli 
spin matrices and the parameter $\theta$ is the rapidity} .

In \cite{MN} the $N=2$ SG model was suggested as a 
good string background. The $N=2$ SG model is known to enjoy
invariance under $N=2$ superalgebra as well as with respect
to the quantum affine algebra $U_{q}(\hat{Sl(2)})$
with $q$ given by (\ref{O}). These two algebras conmute, thus the
whole symmetry of the light cone world sheet model is
\ben\label{S}
U_{q}(Sl(2)) \otimes U_{q^{2}=-1}(Sl(2)) = U_{q}(Sl(2)) \otimes (N=2) 
\een
This invariance is manifest
in the structure of the soliton S-matrices. The S-matrices are 
the product of the standard $N=0$ SG S-matrix
(the $U_{q}(\hat{Sl(2)})$ part that is a dynamical symmetry of $N=0$ SG as
discussed above) and the minimal $N=2$ S-matrix (see
for instance \cite{KU}).
The question we would like to address now is 
what is the stringy meaning of this symmetry.
 
First of all we should notice that this symmetry
will only appear in the string in the deep IR where the string is
effectively very large. The reason is that this quantum symmetry
is corrected by finite size effects. 
 
The symmetry (\ref{S}) should be compared with the
symmetry of the light cone world sheet lagrangian in flat background
where we have also the $N=2$ piece but in addition we have conformal
Virasoro invariance
\ben
Vir \otimes (N=2)
\een

The quantum symmetry
of light cone massive models is not naturally associated with any 
target space time isometry and moreover depends on the value of the
coupling breaking conformal invariance. It is important
to realize that the classical limit $q=1$ does not correspond
to the conformal case. Also it is important to notice that the level
of the quantum deformed Kac Moody algebra is $k=0$ and have nothing
to do with any Kac Moody level we can associate with the
underlying CFT in the UV limit. 
Thus we notice that the massive world sheet light cone lagrangian can
flow in some cases to massive S-matrix theories in the IR. The quantum
affine symmetry fix the S-matrix data in the IR in 
a similar way as conformal invariance
fixes the CFT data in the UV. In the case of LG backgrounds
we have discussed above the world sheet flow goes from a CFT in the UV, the
free field theory, into another CFT in the IR with chiral ring determined by
the superpotential.

\subsection{Beyond pp-waves: Conformal Affine Toda}
Given a massive model as the light cone world sheet lagrangian 
the general procedure
to get a conformal invariant model is to add the two extra fields $X^{+}$ and
$X^{-}$ in such a way that $\partial_{-} \partial_{+} X^{+} = 0$ requiring, at
the same time,
that once we fix the light cone gauge we recover the original massive model.
In simple cases, as we have already
discussed above, the corresponding background metric
is of the pp-wave type. 

From the purely algebraic point of view the promotion of a massive model
possessing a quantum affine symmetries $U_{q}(\hat{Sl(2)})$ -
with $\hat{Sl(2)}$ the affine Kac Moody algebra at level $k=0$ -
to a conformal field theory
requires, in order to have only infinite dimensional irreps as 
it is the case in CFT, 
to add a center to the Kac Moody algebra i.e to go to level $k$
different from zero. The appropiated way to do it
in the case of the massive SG model was first stablished
in reference \cite{BB}. The main idea is to extend the affine Kac Moody
algebra adding the center and the derivation $d$. In this way the generators
of the Cartan subalgebra are, for the case of $\hat{Sl(2)}$
\ben
H, K, d
\een
The Toda field $T$ valued in the Cartan subalgebra
is
\ben
T = \frac{1}{2}\Phi H + \eta d + \frac{1}{2} \epsilon K
\een
and the corresponding lagrangian is
\ben\label{T}
S= \int dx dt ( \frac{1}{2} \partial_{t}\Phi \partial_{t}\Phi
- \frac{1}{2} \partial_{x}\Phi \partial_{x}\Phi + 
\partial_{t}\eta \partial_{t} \epsilon - 
\partial_{x} \eta \partial_{x} \epsilon - 2( e^{2\Phi} + e^{-2\Phi + 2\eta}))
\een
This is in fact a conformal invariant lagrangian that for $ \eta = cte$
reduces to the massive Sinh Gordon model.
Notice also that for (\ref{T}) the equation of motion
of the additional field $\eta$ is
\ben\label{U}
\partial_{+} \partial_{-} \eta = 0
\een
We can interpret the previous algebraic construction as the way to 
promote the massive Sinh Gordon massive model to a CFT just adding
the two extra fields $\eta$ and $\epsilon$. Moreover if we interpret
the two extra fields $\eta$ and $\epsilon$ as the fields $X^{+}$ and $X^{-}$
respectively, we observe that we can take, thanks to (\ref{U}), the 
light cone gauge obtaing the massive Sinh Gordon model in the case $p^{+} = 0$.

In the light cone gauge $\eta = p^{+}t$ we observe that the model
(\ref{T}) flow in time from a Liouville CFT at
$t= - \infty$
\ben
S= \int dx dt ( \frac{1}{2} \partial_{t}\Phi \partial_{t}\Phi
- \frac{1}{2} \partial_{x}\Phi \partial_{x}\Phi - 2( e^{2\Phi})
\een
to a Sinh Gordon at $t=0$ and again to a Liouville CFT at
$t= + \infty$ \footnote{ The case of Sine Gordon can be done using the same
techniques see \cite{BLC}}
\subsection{Holography}
In reference \cite {DGR} it was suggested, in the context of the maximally
symmetric pp-wave, to identify the extra light cone coordinate $X^{+}$
as a sort of holographic coordinate
\footnote{ For other discussions on holography for pp-waves see \cite{O}}
. The previous exercise provides a new
algebraic understanding on the role of this coordinate. Namely after
identifying $\eta$ with $X^{+}$ we observe that the conjugated operator
$d$ is just $ z\frac{d}{dz}$ for $z$ the affine coordinate used in the
definition of the affine Kac Moody algebra. This is just $L_{0}$ or 
in other words the light cone hamiltonian. In this sense
it is natural to interpret the flow described above
between Liouville and Sinh Gordon in holographic terms. 
Notice that in order to make the theory conformal we have added 
extra generators
$K$ and $d$ to the quantum affine symmetry algebra and that in the
``light cone gauge'' defined by $\eta = p^{+}t$ the flow from
conformal to massive is a flow in ``time''.

\section{Final Comments}
In this note we have observed that some string pp-wave backgrounds
define, in the light cone gauge, integrable two dimensional models. One 
particular example is the Sine Gordon model. The massive spectrum in
the infrared, as well
as the integrability structure of this model is determined by the 
quantum affine symmetry generated by non local charges 
we have described above. We can gain some
intuition on this symmetry by considering the discrete version of the
Sine Gordon model that is ,after bosonization, the well known $XXZ$
spin chain. The infinite $XXZ$ chain enjoys the affine quantum symmetry
$U_{q}(\hat sl(2)$. This is non true if we consider the finite chain
where only after introducing appropiated boundary terms we can get
quantum group invariance with respect to $U_{q}(sl(2))$ i.e not affine.
We observe, considering these models, that the quantum affine symmetry
is intimately related with the existence of a mass gap i.e 
with the existence of a mass gap, in the thermodinamic limit, for the
corresponding string light cone hamiltonian. 

From the string theory point
of view the natural question to ask is if for good string backgrounds that
define light cone hamiltonians that flow in the infrared to massive two
dimensional theories, then these massive models are integrable and possissing
quantum affine symmetries. In this note we have suggested that
this is in fact the case. 

After BMN we know that the
spectrum of the light cone hamiltonian
is related , if there exist a gauge holographic dual, with the
spectrum of anomalous dimensions i.e with the spectrum of the dilatation
operator. The quantum symmetries we are refering to commute with
the light cone hamiltonian, and therefore ,if there exist the holographic 
gauge dual, with the dilatation operator.

\section{Acknowledgements} I would like to thank D.Bernard, A. LeClair, 
J. Barbon and E. Lopez for useful discussions. Especially I would like
to thank S.Das for many discussions and initial collaboration on this topic.
I would like also to thank D.Bernard and A. LeClair for pointing out
reference \cite{BB}. This research was supported by grant AEN2000-1584.

\end{document}